\documentclass[aps,preprint,letterpaper,amsmath,amssymb]{revtex4}
\usepackage{graphicx}% Include figure files
\usepackage{dcolumn}% Align table columns on decimal point
\usepackage{bm}
\begin{document}

\title{Angoricity and compactivity describe the jamming transition
  in soft particulate matter}

\author{Kun Wang$^{1}$, Chaoming Song$^{2}$, Ping Wang$^{3}$, Hern\'an A. Makse$^{1}$}

\affiliation {$^{1}$Levich Institute and Physics Department, City College of New York, New York, NY 10031, US \\
  $^{2}$Center for Complex Network Research, Department of Physics, Biology and Computer Science, Northeastern University, Boston, MA 02115, US \\
  $^{3}$FAS Center for Systems Biology, Harvard University, Cambridge,
  MA 02138, US}

\date{\today }

\begin{abstract}
  The application of concepts from equilibrium statistical mechanics
  to out-of-equilibrium systems has a long history of describing
  diverse systems ranging from glasses to granular materials. For
  dissipative jammed systems-- particulate grains or droplets-- a key
  concept is to replace the energy ensemble describing conservative
  systems by the volume-stress ensemble.  Here, we test the
  applicability of the volume-stress ensemble to describe the jamming
  transition by comparing the jammed configurations obtained by
  dynamics with those averaged over the ensemble as a probe of
  ergodicity.  Agreement between both methods suggests the idea of
  ``thermalization'' at a given angoricity and compactivity. We
  elucidate the thermodynamic order of the jamming transition by
  showing the absence of critical fluctuations in static observables
  like pressure and volume. The approach allows to calculate
  observables such as the entropy, volume, pressure, coordination
  number and distribution of forces to characterize the scaling laws
  near the jamming transition from a statistical mechanics viewpoint.
\end{abstract}

\maketitle

A granular system compresses into a mechanically stable configuration
at a nonzero pressure in response to the application of an external
strain \cite{coniglio,behringer,notcool}. This
% due to the dissipative interactions among the constitutive grains, a
process is typically referred to as the jamming transition and occurs
at a critical volume fraction $\rm {\phi_c}$ \cite{notcool}. The
application of a subsequent external pressure with the concomitant
particle rearrangements and compression results in a set of
configurations characterized by the system volume $V=N V_g/\phi$
($\phi$ is the volume fraction of $N$ particles of volume $V_g$) and
applied external stress or pressure $p$ (for simplicity we assume
isotropic states). It has been long argued whether the jamming
transition is a first-order transition at the discontinuity in the
average coordination number, $\langle Z \rangle$, or a second-order
transition with the power-law scaling of the system's pressure as the
system approaches jamming with $\phi-\phi_c \to 0^+$
\cite{jpoint,powerlaw,hmlaw2,mode2}. Previous work
\cite{forcemap,gama,canonicaljam} has proposed to explain the jamming
transition by a field theory in the pressure ensemble. Here, we use
the idea of ``thermalization'' of an ensemble of mechanically stable
granular materials at a given volume and pressure to study the jamming
transition from a thermodynamic viewpoint.

For a fixed number of grains, there exist many jammed states
\cite{manypackings} confined by the external pressure $p$ in a volume
$V$. In an effort to describe the nature of this nonequilibrium system
from a statistical mechanics perspective, a pressure-volume ensemble
\cite{edwardsbook1,angoricity,forcebalance1,forcechain,forcemap} was
introduced for jammed matter. In the canonical the probability of a
state is given by $\exp [-{\cal W} (\partial S/\partial V) - \Gamma
(\partial S/\partial \Gamma)]$, where $S$ is the entropy of the
system, $\cal W$ is the volume function measuring the volume of the
system as a function of the particle coordinates and $\Gamma \equiv
pV$ is the boundary stress (or internal virial) \cite{gama} of the
system. Just as $\partial E/\partial S = T$ is the temperature in
equilibrium system, the temperature-like variables in jammed systems
are the compactivity $X=\partial V/\partial S$ \cite{edwardsbook1} and
the angoricity $A=\partial \Gamma/\partial S$ \cite{angoricity}.

% (from the Greek ``$\acute{\alpha}\gamma\chi o \zeta$'' (ankhos
% =stress)) which are determined by the volume $V$ and the internal
% virial $\Gamma$ (is an extensive quantity) of the system.
In a recent paper \cite{compactivity} the compactivity was used to
describe frictional hard spheres in the volume ensemble. Here, we test
the validity of the statistical approach in the combined
pressure-volume ensemble to describe deformable, frictionless
particles, such as emulsion systems jammed under osmotic pressure near
the jamming transition \cite{forcedist}. We demonstrate that the
jamming transition can be probed thermodynamically by
the angoricity $A$ and the compactivity $X$.
% the divergence We characterize the jamming transition
% thermodynamically by a ``jamming temperature'' $T_{\rm J}$ comprised
% of contributions from Furthermore, the calculation of an effective
% capacity, $C_{T}$ which diverges at the critical volume of the
% system (J-point), demonstrates that jamming transition has a
% second-order transition feature from a thermodynamic point of view.
The calculation of jamming ``heat'' capacities characterizes the
system fluctuations and shows the lack of critical fluctuations in the
static quantities as the jamming transition point is approached from
above $\phi \to \phi_c^+$. Thus, the thermodynamical viewpoint
determines the order of the phase transition and allows one to
calculate the physical observables near jamming.

% Considering a granular system with fixed number grains $n$ and fixed
% volume fraction $\phi$, if Edwards ensemble established, the
% probability of finding this system in one jammed state should obey
% the distribution $\exp (- \alpha p_{i} )$, where $\alpha = 1/A$,
% determined by the pressure $p_{i}$ of this particular jammed
% state. Furthermore, i

\section{Results}

In general, if the density of states $g(\Gamma,\phi)$ in the space of
jammed configurations (defined as the probability of finding a jammed
state at a given $(\Gamma,\phi)$ at $A=\infty$) is known, then
calculations of macroscopic observables, like pressure $p$ and average coordination number $Z$
as a function of $\phi$, can be performed
by the canonical ensemble average \cite{gama,canonicaljam} at a given volume:
% For instance, the ensemble average of $p$ and $Z$ at a fixed volume
% can be realized by:
\begin{equation}
  \langle p(\alpha,\phi) \rangle_{\rm ens} =\frac {1}{\mathcal{Z}}
\int_{0}^{\infty}p\,\,g(\Gamma,\phi)\,\,
    e^{-\alpha \Gamma}\,\,{\rm d}\Gamma,
  \label{Eq_edP}
\end{equation}
and
\begin{equation}
  \langle Z(\alpha,\phi) \rangle _{\rm ens} =\frac {1}{\mathcal Z}
\int_{0}^{\infty}Z \,\,g(\Gamma,\phi)\,\, e^{-\alpha \Gamma}\,\,{\rm d}\Gamma,
\label{Eq_edZ}
\end{equation}
% and
%\begin{equation}
%  \phi_{\rm ens} =\frac {\int_{0}^{\infty} \phi g(p) e^{-\alpha p}{\rm d}p}{\int_{0}^{\infty}g(p) e^{-\alpha p}{\rm %d}p}, \label{Eq_edphi},
%\end{equation}
where the canonical partition function is
$\mathcal{Z}=\int_{0}^{\infty}g(\Gamma,\phi) e^{-\alpha \Gamma}{\rm
  d}\Gamma$ and the density of states is normalized as
$\int_{0}^{\infty}g(\Gamma,\phi){\rm d}\Gamma=1$. The inverse
angoricity is defined as $\alpha \equiv 1/A= \partial S/\partial
\Gamma$.

At the jamming transition the system reaches isostatic equilibrium,
such that the stresses are exactly balanced in the resulting
configuration, and there exists a unique solution to the interparticle
force equations satisfying mechanical equilibrium. It is well known
that observables present power-law scaling
\cite{jpoint,powerlaw,mode2}:
% such as the
%pressure $p$ and the extra number of contacts per particle, $Z-Z_{c}$,
% of the difference between volume fraction $\phi$ and the critical
% volume fraction $\phi_{c}$

\begin{equation}
  \langle p \rangle_{\rm dyn} \sim (\phi-\phi_{c})^{a}\,,
%  \label{Eq_mdP}
%\end{equation}
%\begin{equation}
  \,\,\,\,\,  \,\,\,\,\,  \,\,\,\,\,
 \langle Z \rangle_{\rm dyn}-Z_{c} \sim
  (\phi-\phi_{c})^{b},
  \label{Eq_mdZ}
\end{equation}
where $a=3/2$ and $b=1/2$ for Hertzian spheres and $Z_{c}=6$ is the
coordination number at the isostatic point (J-point) \cite{jpoint}.
The average $\langle \cdots \rangle_{\rm dyn}$ indicates that these
quantities are obtained by averaging over packings generated
dynamically in either simulations or experiments as opposed to the
ensemble average over configurations $\langle \cdots \rangle_{\rm
  ens}$ of Eqs. (\ref{Eq_edP})--(\ref{Eq_edZ}). Comparing the ensemble
calculations, Eq. (\ref{Eq_edP})--(\ref{Eq_edZ}), with the direct
dynamical measurements, Eq.  (\ref{Eq_mdZ}), provides a basic test of
the ergodic hypothesis for the statistical ensemble.

Our approach is the following: We first perform an exhaustive
enumeration of configurations to calculate $g(\Gamma,\phi)$ and obtain
$\langle p(\alpha,\phi) \rangle_{\rm ens}$ as a function of $\alpha$
for a given $\phi$ using Eq. (\ref{Eq_edP}).
% We first perform an ensemble calculation of the pressure From
% Eq. (\ref{Eq_edP}) the ensemble calculation requires the which is
% obtained through an of the jammed states at a fixed volume $V$.
Then, we obtain the angoricity by comparing the pressure in the
ensemble average with the one obtained following the dynamical
evolution with Molecular Dynamics (MD) simulations. By setting
$\langle p(\alpha,\phi) \rangle_{\rm ens} = \langle p \rangle_{\rm
  dyn}$,
% where $\langle p \rangle_{\rm dyn}$ is the MD result, for systems at
% different volume fractions $\phi$, the
we obtain the angoricity as a function of $\phi$.
% using Eq.(\ref{Eq_edP}).
By virtue of obtaining $\alpha(\phi)$, all the other observables can
be calculated in the ensemble formulation.  The ultimate test of
ergodicity is realized by comparing the remaining ensemble
observables with the corresponding direct dynamical
measures.

%$P_{\rm ens}(F/\langle F
%\rangle_{\rm ens}) \rangle$ from the ensemble average with the
% $\langle Z \rangle_{\rm dyn}$ and $\langle P_{\rm dyn}(F/\langle F
% \rangle_{\rm dyn}) \rangle$ using $\alpha$ obtained by
% Eq.(\ref{Eq_edP}).
% the coincidence can provide a good evidence to the establishment of
% Edwards ensemble.

{\bf Ensemble calculations.---} The density of jammed states
$g(\Gamma,\phi)$ is calculated in the framework of the potential
energy landscape (PEL) formulation introduced by Goldstein \cite{PEL}
and Stillinger-Weber \cite{sw1,sw2} to describe supercooled
liquids. In the case of frictionless jammed systems, the mechanically
stable configurations are defined as the local minima of the potential
energy surface (PES) of the system \cite{jpoint,manypackings} (see
Fig. \ref{dos_figure} inset for a schematic representation). In the
simulations, two spherical soft particles in contact interact via a
normal Hertz force \cite{hmlaw1,hmlaw2}, $F_n \propto (\delta r)
^{\delta}$, where $\delta r$ is the normal overlap between the spheres
under deformation and $\delta=1.5$, in a periodically repeated cube
[the interparticle potential energy is $E\propto (\delta
r)^{\delta+1}$, see Materials and Methods Section \ref{hertz}]. The
Hertz potential is chosen for its general applicability to granular
materials.  The results are expected to be independent of the form of
the potential.  Details of the algorithms \cite{LBFGS1,LBFGS2} to find
the local minima of the PES (zero-order saddles) are in the
Materials and Methods Section \ref{dos}.  Figure \ref{dos_figure}
shows $g(\Gamma,\phi)$ versus $\Gamma$ for different volume fractions.
% implying that at low volume fraction the system can not explore the
% complete PEL due to the insufficient degrees of freedom, as compared
% with a system at higher volume fraction.

{\bf MD calculations.---} The pressure $\langle p \rangle_{\rm dyn}$
as a function of $\phi$ is calculated by performing MD
simulations. Packings are prepared by compressing a gas of particles
from an initial (unjammed) low volume fraction to a final jammed
state. This procedure simulates a dynamical packing preparation
\cite{forcebalance2}; details appear in Materials and Methods Section
\ref{MD}.
% From previous studies, it has been observed the pressure $p$
% vanishes as power-law of $\phi$ when approaching the jamming
% transition as seen in Eq.(\ref{Eq_mdP}) \cite{jpoint}.
We obtain (Fig. \ref{ango}A)
% The relation between pressure $\langle p \rangle _{\rm dyn}$ and
% volume fraction $\langle \phi \rangle_{\rm dyn}$ is obtained,
% displaying a typical power-law behavior near the jamming point:
\begin{equation}
  \langle p \rangle _{\rm dyn} = p_{0} \,\,
  ( \phi -\phi_{c})^{1.65} \,\,,
  \label{Eq_mdPvphi}
\end{equation}
where $ \phi_{c}=0.6077$ is the volume fraction corresponding to the
isostatic point J \cite{jpoint} following Eq. (\ref{Eq_mdZ}) and $p_{0}=10.8 {\rm MPa}$.  This
critical value and the exponent, $a = 1.65$, are slightly different than the values
obtained for larger systems ($a=\delta$)\cite{jpoint}. However, our purpose is to
use the same system in the dynamical calculation and the exact
enumeration for a proper comparison.
% By Eq. (\ref{Eq_mdPvphi}) we can thereby calculate the pressure
% $\langle p \rangle_{\rm dyn}$ for any volume fraction $\phi$ near
% $\phi_{c}$ for the particular system under study.

{\bf Calculation of angoricity.---} For each $\phi$ we use
$g(\Gamma,\phi)$ to calculate $\langle p (\alpha)\rangle_{\rm ens}$ by
Eq. (\ref{Eq_edP}). Then, we obtain $\alpha(\phi)$ by setting $\langle
p(\alpha,\phi) \rangle_{\rm ens} = \langle p \rangle_{\rm dyn}$ for
every $\phi$ (see Figs. \ref{alpha_phi} and \ref{p_vs_alpha} and
Materials and Methods Section \ref{angoricity}). The resulting
equation of state $\alpha(\phi)$ is plotted in Fig. \ref{ango}B and
shows that the angoricity follows a power-law, near $\phi_c$, of the
form:
\begin{equation}
A \propto
(\phi-\phi_{c})^{\gamma},
\label{inverse}
\end{equation}
where the angoricity exponent is $\gamma = 2.5$. The result is
consistent with $\gamma = \delta + 1.0$, suggesting that $A \propto
\Gamma \propto F_{n} r$.  Angoricity is a measure of the number of
ways the stress can be distributed in a given volume. Since the
stresses have a unique solution for a given configuration at the
isostatic point, $\phi_c$, the corresponding angoricity vanishes.
%such that the isostatic limit at $\phi_c$
%corresponds to $A=0$.
At higher pressure, the system is determined by multiple degrees of
freedom satisfying mechanical equilibrium, leading to a higher stress
temperature, $A$.
% From Fig. \ref{ango}, we see that $\alpha$ diverges when the volume
% fraction $\phi$ approaches the critical one increases implying that,
% with increasing $\phi$, the angoricity increases, which is predicted
% by the Edwards' ensemble
The angoricity can also be viewed as a scale of stability for the
system at different volume fractions. Systems jammed at larger volume
fractions require higher angoricity (higher driving force) to rearrange.

{\bf Test of ergodicity.---}In principle, using the inverse angoricity, $\alpha$, from
Eq. (\ref{inverse}) we can calculate any macroscopic statistical
observable $\langle B \rangle_{\rm ens}$
at a given volume by performing the ensemble average \cite{canonicaljam}:
\begin{equation}
  \langle B(\phi) \rangle _{\rm ens} =\frac{1} {{\mathcal{Z}}}
  \int_{0}^{\infty}B \,\, g(\Gamma,\phi) \,\, e^{-\alpha \Gamma} \,\,{\rm d}\Gamma.
  \label{Eq_edaB}
\end{equation}
We test the ergodic hypothesis in the Edwards's ensemble
by comparing Eq. (\ref{Eq_edaB}) with
the corresponding value obtained with MD simulations averaged over
% should equal the sample average $\langle B \rangle_{\rm dyn}$.
% where the ensemble average $\langle B \rangle_{\rm ens}$ is
% calculated by the ensemble calculation as: and the sample average
% $\langle B \rangle_{\rm dyn}$ is obtained by averaging of the
($250$) sample packings, $B_i$, generated dynamically:
\begin{equation}
  \langle B(\phi) \rangle _{\rm dyn} =\frac {1}{250}
  \sum_{i=1}^{250} B_{i}.
\label{Eq_mdaB}
\end{equation}

The comparison is realized by measuring the average coordination
number, $\langle Z \rangle$, the average force and the distribution of interparticle
forces. We calculate $\langle Z \rangle_{\rm ens}$ by
Eq. (\ref{Eq_edZ})
% and (\ref{inverse}).
and $\langle Z \rangle_{\rm dyn}$ as in Eq. (\ref{Eq_mdaB}).  Figures
\ref{predict}A and \ref{predict}B show that the two independent
estimations of the coordination number agree very well: $\langle Z
\rangle _{\rm ens}=\langle Z \rangle _{\rm dyn}$.

%for a jammed packing
%is proportional to the pressure of the packing. 
We calculate the ensemble average force $\langle\overline F
\rangle_{\rm ens}$ and the average over all the MD packings,
$\langle\overline F \rangle_{\rm dyn}$ and find that they coincide
very closely (see Fig. \ref{predict}C).  The full distribution of
inter-particle forces for jammed systems is also an important
observable which has been extensively studied in previous works
\cite{jpoint,forcedist1,forcedist2}. The force distribution is
calculated in the ensemble $P_{\rm ens}(F/\overline F)$ by averaging
the force distribution for every configuration in the PES
% , where is the average of $P(F/\langle F \rangle)$ for each
% configuration, where the average force configuration separately,
% that is, just $\overline F \rangle$ is the the average of all the
% forces in one packing configuration
(see Materials and Methods Section \ref{angoricity}). Figure
\ref{predict}D shows the distribution functions. The peak of the
distribution shown in Fig. \ref{predict}D indicates that the systems
are jammed \cite{jpoint,forcedist1,forcedist2}. Besides the exact
shape of the distribution, the similarity between the ensemble and the
dynamical calculations shown in Fig. \ref{predict}D is significant.
The study of $\langle Z \rangle$, $\langle\overline F \rangle$ and
$P(F/\overline F)$ reveals that the statistical ensemble can predict
the macroscopic observables obtained in MD. This suggests that the
idea of ``thermalization'' at an angoricity is able to describe the
jamming system very well.

{\bf Thermodynamic analysis of the jamming transition.---} So far we
have considered how the angoricity determines the pressure
fluctuations in a jammed packing at a fixed $\phi$. The role of the
compactivity in the jamming transition can be analyzed in terms of the
entropy which is easily calculated in the microcanonical ensemble from
the density of states. Figure \ref{entropydis} shows
$S=\ln(\Omega(p,\phi))$ ($\Omega$ is the number of states which is the
unnormalized version of $g(\Gamma,\phi)$), which is the
non-equilibrium entropy of the system at the given $(p,\phi)$ in phase
space.

% superimposed to the equation of state obtained with MD: $\Gamma_{\rm
%   dyn}(\phi)$.  The critical phenomenon of jamming has been observed
% by both experiments and simulations. The behavior of observables,
% such as the contact number $Z$ and the pressure $p$, indicates a
% phase transition at a critical volume fraction. However, such
% different observables give different features of phase
% transition. Here, we characterize the jamming transition from a
% statistical mechanics perspective.  Since $\Gamma_{\rm dyn}$ and
% $\phi$ are coupled by the equation of state coming from the Hertz
% force, the maximum entropy principle states that, Within the
% ensemble calculation, the entropy of the system can be viewed as
% $S(V,\Gamma)=\ln(\Omega(V,\Gamma))$, where $\Omega(V,\Gamma)$ is the
% number of jammed states at a given volume $V$ and $\Gamma$.  At
% equilibrium, the entropy is maximum respect to changes in $V$ and
% $\Gamma$.

We analyze the non-equilibrium entropy surface
$S(\ln(\phi-\phi_{c}),\ln p)$ plotted versus $(\ln(\phi-\phi_{c}),\ln
p)$ in Fig.  \ref{entropydis} and demonstrate that the MD curve
$\langle p(\phi) \rangle_{\rm dyn}$ passes along the maximum of the
entropy surface constrained by the coupling between $p$ and $\phi$,
Eq.  (\ref{Eq_mdZ}). Thus the points along the entropy surface defined
by $\langle p(\phi)\rangle$ correspond to the equilibrium entropy;
such a curve is superimposed to the entropy surface in
Fig. \ref{entropydis}. Due to the coupling through the contact force
law, the maximization of entropy is not on $p$ or $\phi$ alone but on
a combination of both. The entropy $S$ reaches a maximum at the point
$S(\ln(\langle \phi \rangle _{\rm dyn}-\phi_{c}),\ln \langle p \rangle
_{\rm dyn})$ when we move along the direction perpendicular to the
jamming curve $\langle p (\phi) \rangle_{\rm dyn}$ (see the
maximization direction in Fig.  \ref{entropydis}). This is a direct
verification of the second-law of thermodynamics: the dynamical
measures maximize the entropy of the system.
%, $S_{\rm noneq}$.

We can use this result to obtain a relation between angoricity and
compactivity. We write $\ln p = \ln p_0 + a\ln(\phi-\phi_c) $ where
$a$ is the exponent in Eq. (\ref{Eq_mdZ}), such that
$S(\ln(\phi-\phi_c),\ln p)$ is maximized at the MD measures according
to the direction of $(-\sin \theta,\cos \theta)$ ($\tan \theta = a$ is
the slope of the power-law curve in the $\log-\log$ plot in
Fig. \ref{entropydis}). Therefore, neither $A$ nor $X$ can play the
role of the temperature of the system alone, but a combination of both
determined by entropy maximization satisfying the coupling between
stress and strain. Since $\delta S=0$ at $(\ln(\langle \phi \rangle
_{\rm dyn}-\phi_{c}),\ln \langle p \rangle _{\rm dyn})$ along $(-\sin
\theta,\cos \theta)$ then $(\partial S/ \partial \ln p) \cos \theta -
(\partial S/\partial \ln(\phi-\phi_c)) \sin \theta=0$. We obtain $c_1
\alpha + a c_2 \beta = 0$ (where $c_1=\Gamma$ and
$c_2=(\phi-\phi_{c})(N V_{g}/\phi^{2})$) and the relation
% Here, by the calculation of angoricity and the power-law jamming
% curve (see
between $A$ and $X$ (see Fig. \ref{slope} and Materials and Methods
Section \ref{entropy}):

\begin{equation}
 X = - a A (\phi-\phi_{c})/p\phi.
 \label{Eq_AX}
\end{equation}
From Eq. (\ref{Eq_AX}) we obtain that: $X \propto -(\phi-\phi_{c})^{1+a-\gamma}/\phi $
%Considering the works have been
%done before and our system size limit, here we treat $a=2.5$. And
and near $\phi_{c}$:
\begin{equation}
  X \sim -(\phi-\phi_{c})^{2}.
\end{equation}
We notice that the compactivity is negative near the jamming
transition. A negative temperature is a general property of systems
with bounded energy like spins \cite{fluctuation}: the system attains
the larger volume (or magnetization in spins) at $\phi_c$ when $X\to
0^-$ and not $X\to +\infty$ [The bounds $\phi_{c} \le \phi \le 1$
imply that the jamming point at $X\to 0^-$ is ``hotter'' than $X\to
+\infty$. At the same time $A\to 0^+$ since the pressure vanishes].

We conclude that, $A$ and $X$ alone cannot play the role of
temperature. Instead, there is an actual ``jamming temperature''
$T_{\rm J}$ that determines the direction $(-\sin \theta,\cos \theta)$
in the $\log-\log$ plot of
%`real temperature T'', which can determine both $A$ and
%$X$.
Fig. \ref{entropydis} along the jamming equation of state (see
Fig. \ref{slope}).
% we can see such ``real temperature T'' is actually along the jamming
% transition curve.
By maximizing the entropy along this direction we obtain
% We obtain the
$T_{\rm J}$ as a function of $A$ and $X$ (see Materials and Methods
Section \ref{entropy}):
\begin{equation}
  T_{\rm J} =\sin \theta \frac {A}{\Gamma} =\frac{a}{\sqrt{1+a^2}} \frac{A}{\Gamma} \sim (\phi-\phi_c)^{\gamma-a}.
  \label{Eq_T}
\end{equation}

% Since we have two temperature-like variables in jammed systems, $A$
% and $X$.

By the definition of ``heat'' capacity, we obtain two jamming
capacities as the response to changes in $A$ and $X$:
\begin{equation}
  C_{\rm \Gamma} \equiv \partial \Gamma/\partial A \sim (\phi-\phi_{c})^{-1} \sim A^{-2/5}, \,\,\,
  {\rm and}\,\,\,
  C_{\rm V} \equiv \partial V/\partial X \sim (\phi-\phi_{c})^{-1} \sim | X|^{-1/2}.
 \label{Eq_CX}
\end{equation}

From Eq. (\ref{Eq_CX}), the jamming capacities diverge at the jamming
transition as $A \to 0^+$ and $X \to 0^{-}$. However, this result does
not imply that the transition is critical since from Einstein
fluctuation theory applied to pressure and volume \cite{fluctuation}
we obtain (we consider $k_B=1$ for simplicity):
\begin{equation}
  \langle (\Delta \Gamma)^{2} \rangle = A^{2}C_{\rm \Gamma} \sim A^{1.6},\,\,\,\,
  {\rm and} \,\,\,\, \langle (\Delta V)^{2} \rangle = X^{2}C_{\rm V} \sim |X|^{1.5}.
 \label{Eq_DV}
\end{equation}
Thus, the pressure and volume fluctuations near the jamming transition
do not diverge, but instead vanish as $A \rightarrow 0^+$ and $X
\rightarrow 0^{-}$.  From a thermodynamical point of view, the
transition is not of second order due to the lack of critical
fluctuations.  As a consequence, no diverging static correlation
length can be found at the jamming point during isotropic
compression. However, other correlation lengths of dynamic origin may
still exist in the response of the jammed system to perturbations,
such as those imposed by a shear strain or in vibrating modes
\cite{mode1,mode2}. Such a dynamic correlation length would not appear
in a purely thermodynamic static treatment as developed here. We note
though that responses to shear can be treated in the present formalism
by allowing the inverse angoricity to be tensorial
\cite{canonicaljam}.
% characteristic length scale could appear when the system is
% perturbed under shear, as the works have been done by analyze the
% vibrational modes of the Hessian matrix for the system. But here,
% the jamming transition can not be viewed as a second order
% transition and there is no diverging static length scale for our
% static ensembles.
The intensive jamming temperature Eq. (\ref{Eq_T}) gives use to a
jamming effective energy $E_{\rm J}$ as the extensive variable
satisfying $T_{\rm J}=\partial E_{\rm J}/\partial S$ and a full
jamming capacity $C_{\rm J} \sim (\phi-\phi_{c})^{-1}$, which also
diverges at jamming (see Materials and Methods Section
\ref{entropy}). However, the fluctuations of $E_{\rm J}$ defined as
$\langle (\Delta E_{\rm J})^{2} \rangle = T_{\rm J}^{2}C_{\rm J} \sim
T_{\rm J}$ has the same behavior as the fluctuations of volume and
pressure, vanishing at the jamming transition $T_{\rm J}\to 0^+$ [$A
\to 0^+$ in Eq. (\ref{Eq_T})].

\section{Conclusions}

We have suggested that the concept of ``
thermalization '' at a compactivity and angoricity in jammed systems
is reasonable by the direct test of ergodicity. The numerical results
indicate that
%the application of concepts from equilibrium statistical
%mechanics to jammed systems is reasonable and
the full canonical ensemble of pressure and volume describes the
observables near the jamming transition quite well.
%The relation between angoricity $A$ and
%volume fraction $\phi-\phi_{c}$ provides further way to predict the
%properties of the jammed system, such as the capacity $C_{T}$.
From a static thermodynamic viewpoint, the jamming phase transition
does not present critical fluctuations characteristic of second-order
transitions since the fluctuations of several observables vanish
approaching jamming.  The lack of critical fluctuations is respect to
the angoricity and compactivity under isotropic compression in the
jammed phase $\phi \to \phi_{c}^+$, which does not preclude the
existence of critical fluctuations when accounting for the full range
of fluctuations in the liquid to the jammed phase transition from
below $\phi_c$. Thus, a critical diverging length scale might still
appear as $\phi \to \phi_c^-$ \cite{glass0,glass1}.  Our results
suggest an ensemble treatment of the jamming transition. One possible
analytical route to use this formalism would be to incorporate the
coupling between volume and coordination number at the particle level
found in \cite{compactivity} together with similar dependence for the
stress to solve the partition function at the mean field level. This
treatment would allow analytical solutions for the observables with
the goal of characterizing the scaling law near the jamming
transition.

\section{Materials and Methods}
 
\subsection{System Information }
\label{hertz}

The systems used for both, ensemble generation and molecular dynamic
simulation, are the same. They are composed of 30 spherical particles
in a periodic boundary box. The particles have same radius $r = 5 \mu
m$ and interact via a Hertz normal repulsive force without
friction. The interaction is defined as:
\begin{equation}
F_n = \frac{2}{3}~ k_n r^{1/2} (\delta r) ^{\delta},
\end{equation}
where $\delta r= (1/2)[2 r - |\vec{x}_1 -\vec{x}_2|]>0$ is the
normal overlap and $k_n=4 G / (1-\nu)$ is defined in terms of the
shear modulus $G$ and the Poisson's ratio $\nu$ of the material
from which the grains are made and $\delta=3/2$. Here, we use $G=29$ GPa and $\nu =
0.2$ for spherical particles and the density of the particles,
$\rho=2 \times 10^{3}$ kg/m$^3$.

\subsection{Ensemble Generation}
\label{dos}

In this section, we first explain the method to obtain
geometrically distinct minima in the PEL. Then we show that the density of the
states, $g(\Gamma,\phi)$, does not change significantly after sufficient
searching time for the configurations.

For $N$ structureless particles possessing no internal orientational
and vibrational degrees of freedom, at a fixed volume fraction $\phi$,
the potential energy is a $3N-$dimensional function,
$E(r_1,\ldots,r_N)$, depending on the positions $r_i$ of the $N$
particles. In principle, if all local minima corresponding to the
mechanically stable configurations of the PEL are obtained, the
density of states $g(\Gamma,\phi)$ can be calculated.  Such an
exhaustive enumeration of all the jammed states requires that $N$ not
be too large due to computational limits.  On the other hand, in order
to obtain a precise average pressure in the MD simulation, $\langle p
\rangle_{\rm dyn}$, $N$ cannot be too small such that boundary effects
are minimized. Considering these constraints, we choose a $30$
particle system.

In order to enumerate all the jammed states at a given volume fraction
$\phi$, we start by generating initial unjammed packings (not
mechanically stable) performing a Monte Carlo (MC) simulation at a
high, fixed temperature. The MC part of the method applied to the
initial packings assumes a flat exploration of the whole PEL. Every MC
unjammed configuration is in the basin of attraction of a jammed state
which is defined as a local minimum in the PES with a positive
definite Hessian matrix, that is a zero-order saddle. In order to find
such a minimum, we apply the LBFGS algorithm provided by Nocedal and
Liu \cite{LBFGS1}. The procedure is analogous to finding the inherent
structures \cite{LBFGS2} of glassy systems. The LBFGS algorithm is
also similar to the conjugate gradient method employed by O'Hern {\it
  et al.}  \cite{jpoint,manypackings}, but it is computationally more
efficient since it does not require the calculation of the Hessian
matrix of the system at any time step. The PEL for each fixed $\phi$
likely includes millions of geometrically distinct minima by our
simulation results. Therefore, an exhaustive search of configurations
is computationally long. We check that the number of found
configurations has saturated after sufficient trials such that the
density of states $g(\Gamma,\phi)$ has converged to a final shape.

It is also important to determine if the local minima are
distinct. Usually, the eigenvalues of the Hessian matrix at each local
minimum can be used to distinguish these mechanically stable
packings. Here, we follow this idea to compare minima to filter the
symmetric packings. However, instead of calculating the eigenvalues of
each packing, which is very time consuming, we calculate a function of
the distance between any two particles in the packing to improve
search efficiency (for the LBFGS algorithm, we do not need to
calculate the Hessian matrix). For each packing, we assign the
function $Q_{i}$ for each particle in the system:
\begin{equation}
  Q_{i}=\sideset{}{}
  \sum_{1 \le j \le N,\;\ j\neq i} {\tan}^{2}(\frac {\pi r_{ij}^{2}}{3 L^{2}}), \label{Eq_tot}
\end{equation}
where $r_{ij}$ is the distance between particles $i$ and $j$, $L$ is
the system size and $N=30$ is the number of the particles of the
system. We list the $Q_{i}$ for each packing from minimum to maximum
$\lbrace Q_{i}\rbrace (1\le i \le N)$. Since $Q_{i}$ is a higher order
nonlinear function, we can assume that two packings are the same if
they have the same list.  The tolerance is defined as:
\begin{equation}
T =\sqrt {\frac {\sum_{\substack{1\le i \le N}}
(Q_{i}-Q^{'}_{i})^{2}}{N^{2}}},
\end{equation}
where $Q_{i}$ and $Q^{'}_{i}$ are the corresponding values from the
lists of two packings.

\begin{figure}[ht]
  \resizebox{10cm}{!}{\includegraphics{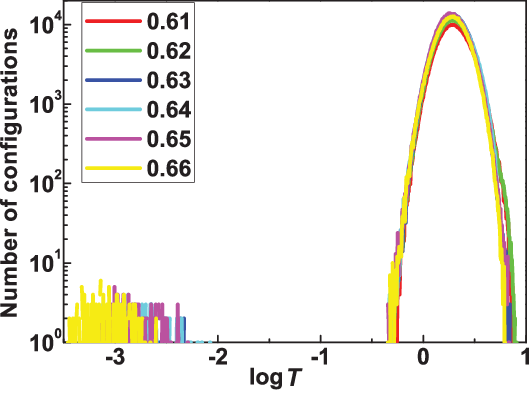}}
  \centering \caption{The distribution of the tolerance $T$ between
    any two packings at the given $\phi$. From the graph, the value of
    $T$ for which any two different packings are considered to be same
    is chosen to be $10^{-1}$, which is above the noise threshold and
    below the distribution of $T$.} \label{tolerance}
\end{figure}

Figure \ref{tolerance} shows the distributions of the tolerance $T$
for packings at different volume fractions. This figure suggests that
two packings can be considered the same if $T \le 10^{-1}$, which
defines the noise level.

\begin{figure}[ht]
  \resizebox{10cm}{!}{\includegraphics{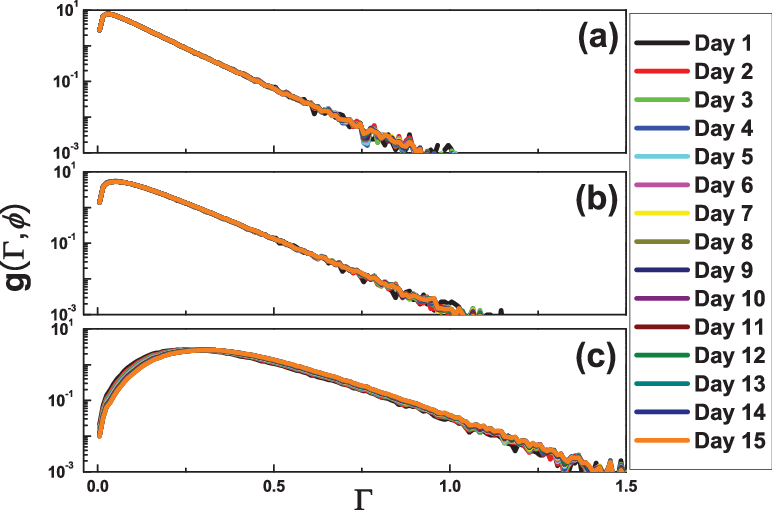}}
  \centering \caption{The distribution of $g(\Gamma,\phi)$ for 15 days
    searching {\it (A)} at $\phi=0.609$, {\it (B)} at $\phi=0.614$,
    {\it (C)} at $\phi=0.625$. Different color in {\it (A)}, {\it
      (B)}, {\it (C)} corresponds to the different day. We find that
    after 15 days the distributions have converged.} \label{dos_phi}
\end{figure}

\begin{figure}[ht]
  \resizebox{10cm}{!}{\includegraphics{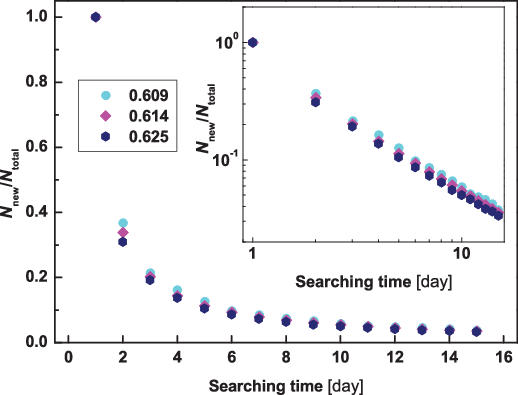}}
  \centering \caption{The probability to find new configurations for
    different searching day.} \label{number}
\end{figure}

From Fig. \ref{dos_phi}, we see that after one week of searching,
$g(\Gamma,\phi)$ does not change significantly, since the initial
packings are generated by a completely random protocol.  We also check
the probability (defined as $\frac {N_{\rm new}(i)}{N_{\rm
    total}(i)}$, where $N_{\rm new}(i)$ is the number of new
configurations found on the $i$th day and $N_{\rm total}(i)$ is the
total number of configurations found in $i$ days) of finding new
mechanically stable states for different searching days. From
Fig. \ref{number}, we see that, after one week searching, the
probability of finding new configurations at different volume
fractions converges, suggesting that enough ensemble packings have
been obtained to capture the features of $g(\Gamma,\phi)$. A further
test of convergence is obtained below in Fig. \ref{checkalpha}.

\subsection{MD Generation}
\label{MD}

In order to analyze numerical results, we perform MD simulations to
obtain $Z_{\rm dyn}$, $p_{\rm dyn}$ and $\phi_{\rm dyn}$, which are
herein considered real dynamics. The algorithm is described in detail
in \cite{hmlaw2,compactivity,forcebalance2}.  Here, a general
description is given: A gas of non-interacting particles at an initial
volume fraction is generated in a periodically repeated cubic
box. Then, an extremely slow isotropic compression is applied to the
system. The compression rate is $\Gamma_0=5.9 t_0^{-1}$, where the
time is in units of $t_0=R\sqrt{\rho/G}$. After obtaining a state for
which the pressure $p$ is a slightly higher than the prefixed pressure
we choose, the compression is stopped and the system is allowed to
relax to mechanical equilibrium following Newton's equations. Then the
system is compressed and relaxed repeatedly until the system can be
mechanically stable at the predetermined pressure. To obtain the
statistical average of $Z_{\rm dyn}$ and $\phi_{\rm dyn}$, we repeat
the simulation to get enough packing samples having statistically
independent random initial particle positions. Here, 250 independent
packings are obtained for each fixed pressure (see
Fig. \ref{all_md}). 
% $\phi=\langle \phi \rangle_{\rm dyn}$ and $\langle Z \rangle_{\rm
%   dyn}$ are flat averages of these $250$ packings by $\langle \phi
% \rangle_{\rm dyn}=\frac {\sum_{1 \le i \le 250} \phi_{i}}{250}$ and
% $\langle Z \rangle_{\rm dyn}=\frac {\sum_{1 \le i \le 250}
%   Z_{i}}{250}$.

\subsection{Angoricity Calculation}
\label{angoricity}

Since we obtain $g(\Gamma,\phi)$ and $\langle p \rangle_{\rm dyn}$ for
each volume fraction $\phi$, we can calculate the inverse angoricity
$\alpha$ by Eq. (\ref{Eq_edP}). The pressure $\langle p(\alpha,\phi)
\rangle_{\rm ens}$ for a given $\phi$ is a function depending on $\alpha$ as:

\begin{equation}
  \langle p(\alpha,\phi)\rangle_{\rm ens} =\frac {\int_{0}^{\infty}pg(\Gamma,\phi)
    e^{-\alpha \Gamma}{\rm d}\Gamma}{\int_{0}^{\infty}g(\Gamma,\phi)
    e^{-\alpha \Gamma}{\rm d}\Gamma}
    =\frac{\sum p e^{-\alpha \Gamma}}{\sum e^{-\alpha \Gamma}}.
  \label{Eq_sedP}
\end{equation}

\begin{figure} [ht]
  \resizebox{10cm}{!}{\includegraphics{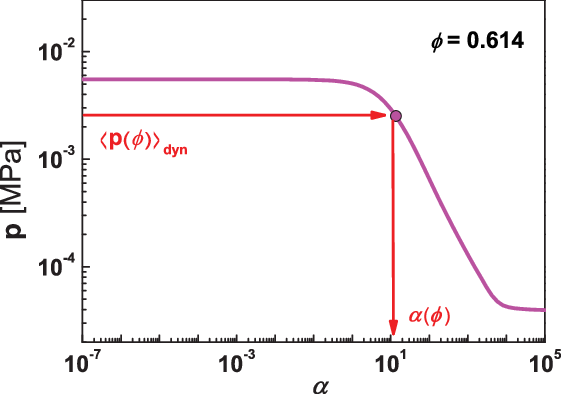}}
  \centering \caption{The numerical integration of Eq. (\ref{Eq_sedP})
    for $\phi=0.614$ is shown as the pink curve. We input the $\langle
    p \rangle_{\rm dyn}$ (pink $\bigcirc$ in the plot) and obtain the
    corresponding inverse angoricity $\alpha$.} \label{alpha_phi}
\end{figure}

\begin{figure} [ht]
  \resizebox{10cm}{!}{\includegraphics{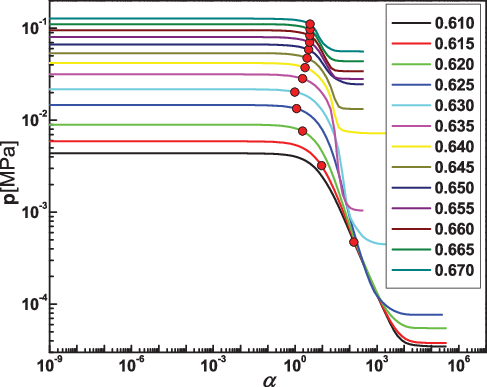}}
  \centering \caption{Calculation of $\alpha$ for several volume
    fractions $\phi$ as explained in detail in
    Fig. \ref{alpha_phi}} \label{p_vs_alpha}
\end{figure}

Figure \ref{alpha_phi} shows the result of the numerical
integration of Eq. (\ref{Eq_sedP}) for a particular $\phi=0.614$ as a
function of $\alpha$ using the numerically obtained $g(\Gamma,\phi)$
from Fig. \ref{dos_figure}. To obtain the value of $\alpha$ for this
$\phi$, we input the corresponding measure of the pressure obtained
dynamically $\langle p(\phi) \rangle_{\rm dyn}$ and obtain the value
of $\alpha$ as schematically depicted in Fig. \ref{alpha_phi}.  The
same procedure is followed for every $\phi$ (see
Fig. \ref{p_vs_alpha}) and the dependence $\alpha(\phi)$ is
obtained. The result is shown in Fig. \ref{ango}B in the main text.

We also check the inverse angoricity $\alpha(\phi)$ using
$g(\Gamma,\phi)$ for different searching days (see Fig. \ref{dos_phi})
to ensure the accuracy and convergence to the proper value. From
Fig. \ref{checkalpha}, we can see that, after 10 days searching,
$\alpha(\phi)$ is stable due to the fact that the density of state,
$g(\Gamma,\phi)$, does not change significantly. For volume fraction
much larger than $\phi_c$, the system's input pressure $\langle
p(\phi) \rangle_{\rm dyn}$ reaches the plateau at low $\alpha$ of the
function $\langle p(\alpha,\phi)\rangle_{\rm ens}$ (see
Fig. \ref{p_vs_alpha}) and the corresponding $\alpha(\phi)$ becomes
much smaller (the angoricity $A(\phi)$ becomes much larger), leading
to large errors in the value of $A$ as $\phi$ becomes large. This
might explain the plateau found in $A$ when $(\phi - \phi_c) > 2
\times 10^{-2}$ as shown in Fig. \ref{ango}B.

Using $\alpha(\phi)$ for each volume fraction, we calculate $\langle Z
\rangle_{\rm ens}$ by:
\begin{equation}
  \langle Z(\phi) \rangle_{\rm ens} =\frac {\int_{0}^{\infty}Zg(\Gamma,\phi)
    e^{-\alpha \Gamma}{\rm d}\Gamma}{\int_{0}^{\infty}g(\Gamma,\phi)
    e^{-\alpha \Gamma}{\rm d}\Gamma}
  =\frac{\sum Z e^{-\alpha \Gamma}}{\sum e^{-\alpha \Gamma}},
  \label{Eq_sedZ}
\end{equation}
the average force $\langle\overline F \rangle_{\rm ens}$ by:

\begin{equation}
  \langle\overline F(\phi) \rangle_{\rm ens} =\frac {\int_{0}^{\infty} \overline F g(\Gamma,\phi)
    e^{-\alpha \Gamma}{\rm d}\Gamma}{\int_{0}^{\infty}g(\Gamma,\phi)
    e^{-\alpha \Gamma}{\rm d}\Gamma}
  =\frac{\sum \overline F e^{-\alpha \Gamma}}{\sum e^{-\alpha \Gamma}},
  \label{Eq_sedF}
\end{equation}
where $\overline F$ is the average force for each ensemble packing and
the force distribution $P_{\rm ens}(F/\overline F)$ by:
\begin{equation}
   P _{\rm ens}(F/\overline F) =\frac {\int_{0}^{\infty}P(F/\overline F )g(\Gamma,\phi) e^{-\alpha \Gamma}{\rm d}\Gamma}{\int_{0}^{\infty}g(\Gamma,\phi)
    e^{-\alpha \Gamma}{\rm d}\Gamma}
    =\frac{\sum P(F/\overline F ) e^{-\alpha \Gamma}}{\sum e^{-\alpha \Gamma}}.
    \label{Eq_sedPro}
\end{equation}
Equations (\ref{Eq_sedZ})--(\ref{Eq_sedPro}) are then compared with
the dynamical measures for a test of ergodicity in Fig. \ref{predict}
in the main text.

\subsection{Entropy Calculation}
\label{entropy}

\begin{figure} [h]
  \resizebox{14cm}{!}{\includegraphics{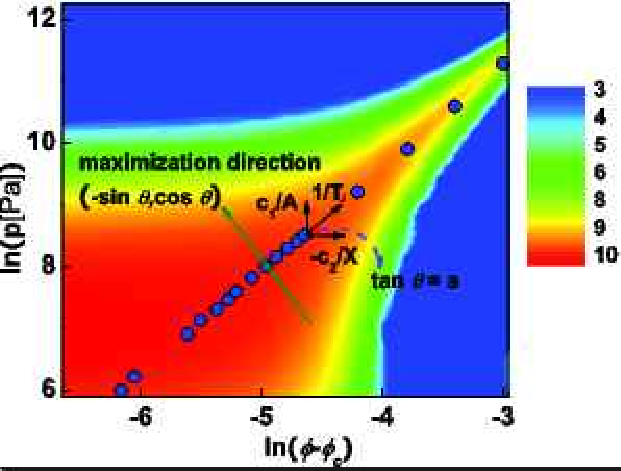}}
  \centering \caption{} \label{entropydis}
\end{figure}

\begin{figure} [ht]
  \resizebox{9cm}{!}{\includegraphics{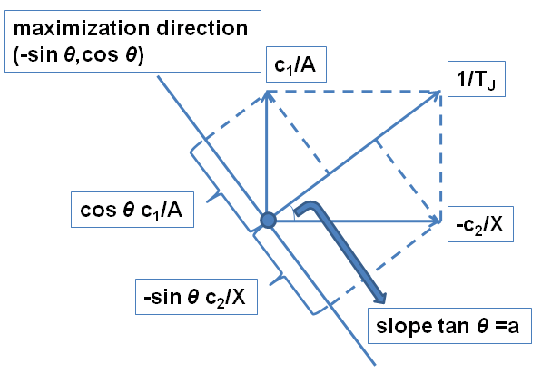}}
  \centering \caption{The representation of the maximization analysis
    $\delta S=0$ along the direction $(-\sin \theta, \cos \theta)$ for
    one point in the jamming power-law curve. Here $c_1=\Gamma$ and
    $c_2=(\phi-\phi_{c})(N V_{g}/\phi^{2})$.}
   \label{slope}
\end{figure}

Here we present the calculation of the ``jamming temperature'' $T_{\rm
  J}$ and the corresponding jamming ``heat'' capacity $C_{\rm
  J}$. From the power-law relation $p=\Gamma/V
\propto(\phi-\phi_{c})^{a}$, we have:
\begin{equation}
 \ln p = \ln p_0 + a\ln(\phi-\phi_c), \label{Eq_powerlog0}
\end{equation}
where $p_0$ is the constant depending on the system and the slope
$\tan \theta = a$. Figure \ref{entropydis} indicates that the
jammed system always remain at the positions of maximal entropy
$\delta S=0$ in the direction ($-\sin \theta$,$\cos \theta$),
perpendicular to the jamming power-law curve. In order to further
analyze this result, we plot the entropy distribution along the
direction ($-\sin \theta$,$\cos \theta$) in Fig.
\ref{distentropy}. We see that the entropy of the corresponding jammed
states remains at the peak of the distributions along ($-\sin
\theta$,$\cos \theta$), verifying the maximum entropy principle in
this particular direction.  We notice that some deviations are found
in the vicinity of $\phi_c$.  The maximization of entropy is not on
$\Gamma$ or $V$ alone, but on a combination of both.  This means that
the entropy $S(\ln(\langle \phi \rangle _{\rm dyn}-\phi_{c}),\ln
\langle p \rangle _{\rm dyn})$ is maximum along the direction of
($-\sin \theta$,$\cos \theta$) and the slope of the entropy along this
direction ($-\sin \theta$,$\cos \theta$) is $0$ (see Fig.
\ref{slope}), that is,
\begin{equation}
 \frac {\partial S}{\partial \ln(\phi-\phi_{c})}\sin \theta = \frac {\partial S}{\partial \ln p}\cos \theta.
  \label{Eq_powerlog1}
\end{equation}
Thus we verify the second law of thermodynamics for jammed systems:
$\delta S=0$ at $(\ln(\langle \phi \rangle _{\rm dyn}-\phi_{c}),\ln
\langle p \rangle _{\rm dyn})$.

By the definition of angoricity $A=\partial \Gamma/\partial S$ and
compactivity $X=\partial V/\partial S$, we have:
\begin{equation}
 \frac {\partial S}{\partial \ln p} = p \frac {\partial S}{\partial p} = \Gamma \frac {\partial S}{\partial \Gamma} = \frac {\Gamma}{A} = \frac {c_1}{A},
 \label{Eq_powerlog2}
\end{equation}

\begin{equation}
  \frac {\partial S}{\partial \ln(\phi-\phi_{c})} = (\phi-\phi_{c}) \frac {\partial S}{\partial \phi}
  = (\phi-\phi_{c})\frac {\partial V}{\partial \phi}\frac {1}{X}=-(\phi-\phi_{c})\frac {N V_{g}}{\phi^{2}}\frac{1}{X}=-\frac {c_2}{X},
  \label{Eq_powerlog3}
\end{equation}
where $\phi=N V_{g}/V$, $c_1=\Gamma$ and $c_2=(\phi-\phi_{c})(N
V_{g}/\phi^{2})$.  By Eq. (\ref{Eq_powerlog2}) and
Eq. (\ref{Eq_powerlog3}), we can simplify Eq. (\ref{Eq_powerlog1}):
\begin{equation}
 \frac{c_1}{A}  + a \frac{c_2}{X} = 0.
  \label{Eq_powerlog4}
\end{equation}

The relation between $X$ and $A$ can be obtained then:
\begin{equation}
  X = -a \frac{c_2} {c_1} A=- a \frac{\phi-\phi_{c}}{p\phi}A.
  \label{Eq_powerlog5}
\end{equation}

Since we obtain the angoricity $A \propto (\phi-\phi_{c})^{\gamma}$
with $\gamma = 2.5$ in the main text and the pressure $p \propto
(\phi-\phi_{c})^{a}$ with $a = 1.5$ (actually we get 1.65 for the
small system size used in the main text but the difference can be
neglected to simplify). The compactivity $X \propto
-(\phi-\phi_{c})^{2}/\phi$. We can therefore define the ``jamming
temperature'' $T_{\rm J}$ as a function of the slope along the
direction ($\cos \theta$,$\sin \theta$):
\begin{equation}
  \frac {1}{T_{\rm J}} = \frac{c_1}{A}\sin \theta - \frac{c_2}{X} \cos \theta = \cos \theta (a \frac{c_1}{A} - \frac{c_2}{X}) = \frac{c_1}{A \sin \theta}=-\frac{c_2}{X \cos \theta}.
  \label{Eq_powerlog6}
\end{equation}
That is:
\begin{equation}
  T_{\rm J} = \frac{A \sin \theta}{c_1}=-\frac{X \cos \theta}{c_2} = \frac {\sin \theta}{\Gamma}A=\frac{a}{\sqrt{1+a^2}} \frac{A}{\Gamma} \sim (\phi-\phi_{c})^{\gamma-a} \sim (\phi-\phi_{c}).
  \label{Eq_powerlog7}
\end{equation}

Furthermore, the ``jamming energy'' $E_{\rm J}$, corresponding to the
``jamming temperature'' $T_{\rm J}$ in Eq.  (\ref{Eq_powerlog7}), has
the relation as below:
\begin{equation}
\begin{split}
  {\rm d}E_{\rm J} &=  T_{\rm J}{\rm d}S \\
  &= T_{\rm J} \frac{\partial S}{\partial \ln(\phi-\phi_c)} {\rm d}\ln(\phi-\phi_c) + T_{\rm J} \frac{\partial S}{\partial \ln p} {\rm d}\ln p \\
  &= (-\frac{X \cos \theta}{c_2})(-\frac {c_2}{X}) {\rm d}\ln(\phi-\phi_c) + \frac{A \sin \theta}{c_1} \frac {c_1}{A}{\rm d}\ln p \\
  &= \cos \theta {\rm d}\ln(\phi-\phi_c) + \sin \theta {\rm d}\ln p \\
  &=(\cos \theta + \sin \theta \tan \theta) {\rm d}\ln(\phi-\phi_c) \\
  &=\frac {{\rm d}\ln(\phi-\phi_c)}{\cos \theta}.
  \label{Eq_powerlog8}
\end{split}
\end{equation}
That is,
\begin{equation}
 {\rm d}E_{\rm J} = \sqrt {a^2+1} {\rm d}\ln (\phi-\phi_c),
  \label{Eq_EJ2}
\end{equation}
and
\begin{equation}
 E_{\rm J} = (\sqrt {a^2+1}) \ln (\phi-\phi_c).
  \label{Eq_EJ3}
\end{equation}

The jamming capacity $C_{\rm J}$ can be obtained as:
\begin{equation}
  C_{\rm J} = T_{\rm J} \frac{\partial S}{\partial T_{\rm J}} = T_{\rm J}\frac{\partial S}{\partial \ln p}\frac{\partial \ln p}{\partial T_{\rm J}} + T_{\rm J}\frac{\partial S}{\partial \ln (\phi-\phi_{c})} \frac{\partial \ln (\phi-\phi_{c})}{\partial T_{\rm J}},
  \label{Eq_powerlog9}
\end{equation}
Finally, with Eq. (\ref{Eq_powerlog1})--(\ref{Eq_powerlog3}), the capacity $C_{\rm J}$ can be calculated:
\begin{equation}
  C_{\rm J} = T_{\rm J} (\frac{c_1}{A}-\frac{c_2}{aX})\frac{\partial \ln p}{\partial T_{\rm J}} = T_{\rm J}\frac{1+a^2}{a^2}\frac{c_1}{A}\frac{\partial \ln p}{\partial T_{\rm J}}.
  \label{Eq_powerlog10}
\end{equation}
Since $T_{\rm J} \sim (\phi-\phi_{c})$ and $p \sim (\phi-\phi_{c})^{1.5}$,
we obtain $C_{\rm J} \sim (\phi-\phi_{c})^{-1}$.

\clearpage

\begin{figure}
  \resizebox{14cm}{!}{\includegraphics{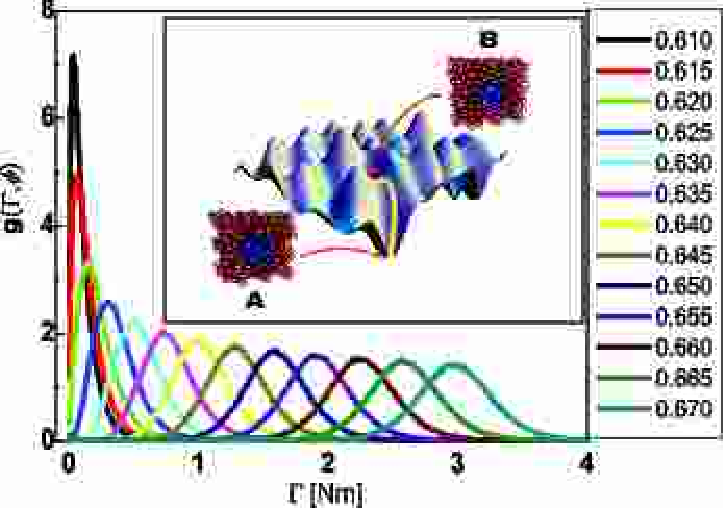}}
  \centering \caption{} \label{dos_figure}
\end{figure}

Fig. \ref{dos_figure}. Ensemble calculations. The density of states
    $g(\Gamma,\phi)$ as a function of internal virial $\Gamma$ for
    different volume fraction, $\phi$, ranging from 0.610 to
    0.670. The inset is a schematic two-dimensional potential energy landscape
  surface. The jammed states $A$ and $B$ are local minima (zero order
  saddles) in the PES where the external force for each particle is
  zero and the Hessian matrix of the system is positive definite. Our
  simulation system consists of $30$ frictionless spherical particles
  interacting by Hertzian forces with periodic boundary conditions.

\begin{figure} [h]
  \resizebox{14cm}{!}{\includegraphics{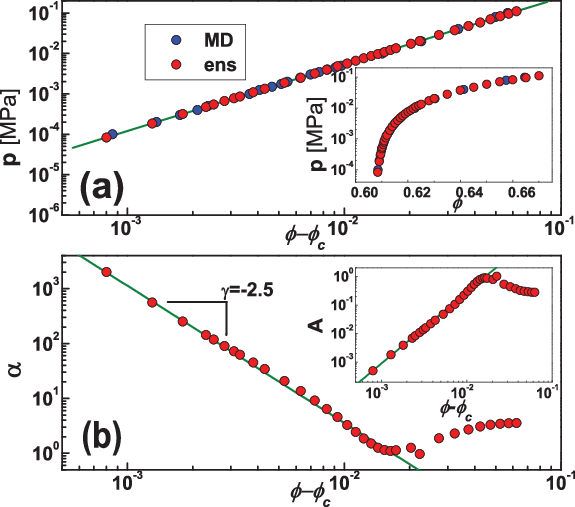}}
  \centering
  \caption{} \label{ango}
\end{figure}

  Fig. \ref{ango}.  Scaling of pressure and inverse angoricity.  {\it
    (A)} The blue $\bigcirc$ shows the power-law relation for $\langle
  p \rangle_{\rm dyn}$ vs $\langle \phi \rangle_{\rm dyn}- \phi_{c} $
  for the 30-particle system. Here, the pressure $\langle p
  \rangle_{\rm dyn}$ are average values obtained by 250 independent MD
  simulations. The red $\bigcirc$ is the pressure used to obtain the
  inverse angoricity $\alpha$ predicted by Eq. (\ref{Eq_mdPvphi}). The
  relatively small system size results in large fluctuations of the
  observables. In order to predict a precise relation for the system
  ($N=30$), sufficient independent samples of the packings are
  generated to calculate the precise average for observables. We
  prepare 250 independent packings for each $\phi$ to get enough
  statistical samples to obtain $\langle p \rangle_{\rm dyn}$ and
  $\langle Z \rangle_{\rm dyn}$ by statistical average (see
  Fig. \ref{all_md}). The inset shows a semi-log plot.  {\it (B)}
  The inverse angoricity $\alpha$ as a function of
  $\phi$-$\phi_{c}$. We find a power-law relation for system's volume
  fraction $\phi$ near $\phi_{c}$. The solid line has a slope of
  -2.5. The inset is the angoricity $A(=1/\alpha)$ vs
  $\phi$-$\phi_{c}$.
  % To find $A$ accurately for system's volume fraction $\phi$ much
  % larger than $\phi_{c}$, becomes difficult due to the large
  % fluctuations and finite size effects. In principle, we
  The plateau observed in $A$ for large volume fraction $\phi$ might
  be related to the finite size of the sample.

\begin{figure} [h]
  \resizebox{14cm}{!}{\includegraphics{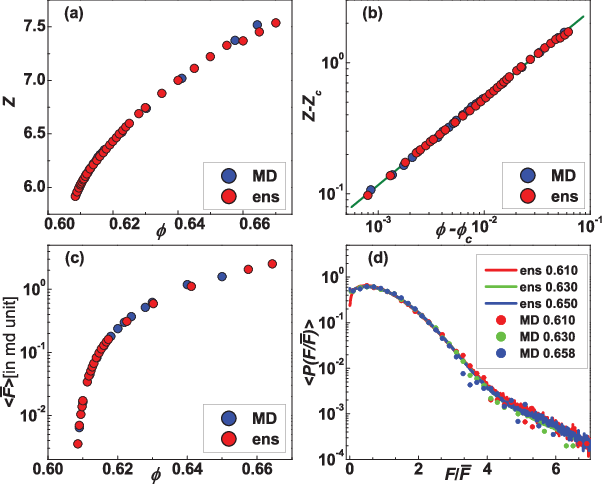}}
  \centering \caption{} \label{predict}
\end{figure}

  Fig. \ref{predict}. Test of ergodicity.  {\it (A) } The blue
  $\bigcirc$ is the average coordination number $\langle Z \rangle
  _{\rm dyn}$ obtained by 250 independent MD simulations. The red
  $\bigcirc$ is the coordination number $\langle Z \rangle _{\rm ens}$
  calculated by the ensemble for different volume fractions. Agreement
  between both measures supports the concept of ergodicity in the
  system.  {\it (B)} The same as {\it (A)} but in a log-log plot. The
  blue $\bigcirc$ shows the power-law relations for $\langle Z
  \rangle_{\rm dyn}$-$Z_{c}$ vs $\langle \phi \rangle_{\rm dyn}$
  -$\phi_{c}$ for 30-particle system with $\phi_{c}=0.6077$ and
  $Z_{c}=5.82$. {\it (C)} Comparison of $\langle \overline F
  \rangle_{\rm dyn}$ and $\langle \overline F\rangle_{\rm ens}$ for
  different volume fractions.  {\it (D)} The comparison of selected
  distributions of forces $ P _{\rm dyn}(F/\overline F )$ and $ P_{\rm
    ens}(F/\overline F ) $ for different volume fractions.

  Fig. \ref{entropydis}.  Microcanonical calculations.  The
  entropy surface $S(\ln(\phi-\phi_{c}),\ln p)$. The color bar
  indicates the value of the entropy. The superimposed blue $\bigcirc$
  is $\langle p(\phi) \rangle_{\rm dyn}$ from MD calculations as in
  Fig. \ref{ango}a. The olive arrow line indicates the maximization
  direction of the entropy $(-\sin \theta,\cos \theta)$. Following
  this direction, the entropy is maximum at the point $(\ln(\langle
  \phi \rangle _{\rm dyn}-\phi_{c}),\ln \langle p \rangle _{\rm
    dyn})$, corroborating the maximum entropy principle.

\clearpage

\clearpage

% We find that $g(\Gamma,\phi)$ does not change significantly after
% sufficient searching time since the initial packings are generated
% by a completely random protocol.

%\section{MD Generation}
%\label{MD-si}

\begin{figure}[ht]
%  \resizebox{10cm}{!}{\includegraphics{nall_md.eps}}
  \centering \caption{The cyan $\bigcirc$ is $Z_{\rm dyn}$ and
    $\phi_{\rm dyn}$ for every single packing obtained with MD and the
    blue $\bigcirc$ are the average over the single packings for the
    system which are then shown in the main text of the
    paper. } \label{all_md}
\end{figure}

%\section{Angoricity Calculation}
%\label{angoricity-si}

\begin{figure} [ht]
  \resizebox{10cm}{!}{\includegraphics{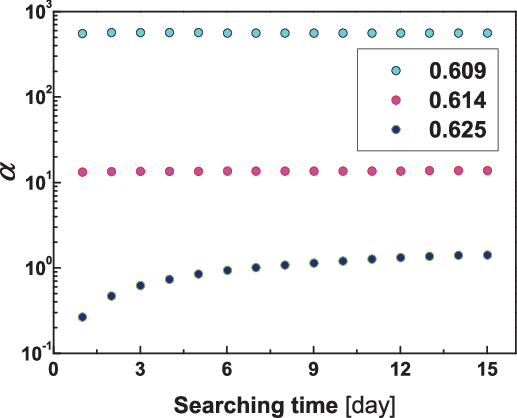}}
  \centering \caption{Calculation of inverse angoricity $\alpha$ for
    different searching day.} \label{checkalpha}
\end{figure}

%\section{Entropy Calculation}
%\label{entropy-si}

\begin{figure} [ht]
%  \resizebox{10cm}{!}{\includegraphics{ndistax.eps}}
  \centering \caption{The distribution of entropy $S(\ln p,
    \ln(\phi-\phi_{c}))$ along the direction $(-\sin \theta,\cos
    \theta)$ for different jamming ensemble points. The blue
    $\bigcirc$ are the entropy for jammed system, which is the maximum
    of $S$, verifying the second law of thermodynamics. We notice that
    some deviations are found near $\phi_c$.}
   \label{distentropy}
\end{figure}

\clearpage

\end{document}